\documentclass[10pt,aps,prl,preprintnumbers,superscriptaddress,showpacs,
twocolumn,floatfix,nofootinbib]{revtex4-1}
\usepackage{graphicx}
\usepackage{amsmath}
\usepackage{slashed}

\newcommand{\bfm}[1]{\mbox{\boldmath$#1$}}

\newcommand{\gsim}{\;\rlap{\lower 3.5 pt \hbox{$\mathchar \sim$}} \raise 1pt
\hbox {$>$}\;}
\newcommand{\lsim}{\;\rlap{\lower 3.5 pt \hbox{$\mathchar \sim$}} \raise 1pt
\hbox {$<$}\;}

\begin{document}

\title{
\boldmath High-Energy Limit of QCD beyond Sudakov
Approximation
\unboldmath}
\author{Tao Liu}
\affiliation{Department of Physics, University of Alberta, Edmonton, Alberta T6G
2J1, Canada}
\email[]{ltao@ualberta.ca}
\author{Alexander A. Penin}
\affiliation{Department of Physics, University of Alberta, Edmonton, Alberta T6G
2J1, Canada}
\affiliation{Institut f\"ur Theoretische Teilchenphysik,
 Karlsruher Institut f\"ur Technologie (KIT), 76128 Karlsruhe, Germany}
 \email[]{penin@ualberta.ca}
\begin{abstract}
We study the high-energy limit of the scattering amplitudes  suppressed by
the leading power of the quark mass in perturbative  quantum chromodynamics.
We prove the factorization and perform  all-order  resummation of the
double-logarithmic  radiative corrections which determine the asymptotic
behavior of the amplitudes.  In contrast to the  Sudakov logarithms, the
mass-suppressed double-logarithmic corrections are induced by soft quark
exchange. The structure of the corrections and the asymptotic behavior of the
amplitudes in this case crucially depend on the color flow in a given process
and are determined by the eikonal color charge nonconservation.  We present
explicit results for the Higgs boson production in gluon fusion mediated by a
light-quark loop and for the  leading power-suppressed contributions  to the
quark form factors, which reveal   ``magical'' universality. Nontrivial
relations between the asymptotic behavior of different amplitudes and
the amplitudes in different gauge theories are found.
\end{abstract}
\preprint{ALBERTA-THY-8-17}

\maketitle
Evaluating the scattering amplitudes  in the high-energy limit is a
fundamental problem of  quantum field theory which remains in the focus of
theoretical research for decades  since the leading asymptotic behavior of an
electron scattering amplitude in quantum electrodynamics (QED) has been
derived in Ref.~\cite{Sudakov:1954sw}.  This behavior is determined by the
``Sudakov'' radiative corrections, which include the second power of the
large logarithm of the electron mass  divided by a characteristic momentum
transfer of the process per each power of the fine structure constant.
Sudakov logarithms exponentiate and result in a strong universal suppression
of the scattering amplitudes in the limit when all the kinematic invariants
of the process are large. Within different approaches the analysis has been
extended to  nonabelian gauge theories and to subleading logarithms
\cite{Frenkel:1976bj,Mueller:1979ih,Collins:1980ih,Sen:1981sd,Sterman:1986aj},
which is crucial for  a wide class of applications  from deep inelastic
scattering to Drell-Yan processes and the Higgs boson production. This
analysis however does not extend to the part of the amplitudes which is power
suppressed in the high energy limit. The power-suppressed  contributions now
attract  a lot of attention  in various contexts (see {\it e.g.}
\cite{Ferroglia:2009ep,Banfi:2013eda,Becher:2013iya,deFlorian:2014vta,
Anastasiou:2014lda,Penin:2014msa,Melnikov:2016emg,Penin:2016wiw,
Bonocore:2016awd,Boughezal:2016zws,Moult:2017rpl,Liu:2017axv}).
Incorporating the logarithmically enhanced power-suppressed terms can
significantly increase the accuracy and extend the region where  the
leading-power approximation is applicable. It becomes crucial when a
power-suppressed term gives the leading contribution to a physical observable
as in the case of the bottom-quark mediated Higgs boson production
in gluon fusion. The latter is formally suppressed by the ratio $m_b^2/m_H^2$
of the quark mass $m_b$ to the Higgs boson mass $m_H$  but
significantly changes the shape of the Higgs boson transverse  momentum
distribution. The effective expansion parameter
in this case is $\ln^2\!(m_b^2/m_H^2)\alpha_s \approx 40 \alpha_s$ rather
than the strong coupling constant  $\alpha_s$, and the resummation of the
double-logarithmic corrections is  mandatory for a reliable
theoretical prediction \cite{Banfi:2013eda,Melnikov:2016emg}. In general very
little is known  so far about the  all-order structure of such corrections.
In contrast to Sudakov logarithms they  do not exponentiate and do not
factorize into the wave functions of scattering particle.   A few known
examples of  the all-order resummation are restricted to  abelian gauge
theory \cite{Penin:2014msa,Melnikov:2016emg,Gorshkov:1966ht,Kotsky:1997rq}.
Extension of the analysis to quantum chromodynamics (QCD) is not
straightforward and requires a  systematic treatment of the factorization. In
this Letter we present the factorization formula  and perform the resummation
of  the double-logarithmic corrections to the amplitudes suppressed by the
leading power of  quark mass.  The Letter is organized as follows. First we
briefly describe the idea of the method  with  a  simple example which
retains all the features of the general problem. Then we apply the resulting
factorization formula to derive the double-logarithmic approximation for the
amplitude of the Higgs boson production in gluon fusion mediated by a
light-quark loop and for the leading power-suppressed contribution to  the
quark scattering in the external vector, axial, scalar and pseudoscalar
fields.  Finally we summarize the qualitative features and discuss
universality of our solution for different amplitudes and gauge models.

\begin{figure}[t]
\begin{center}
\begin{tabular}{cccc}
\includegraphics[width=1.5cm]{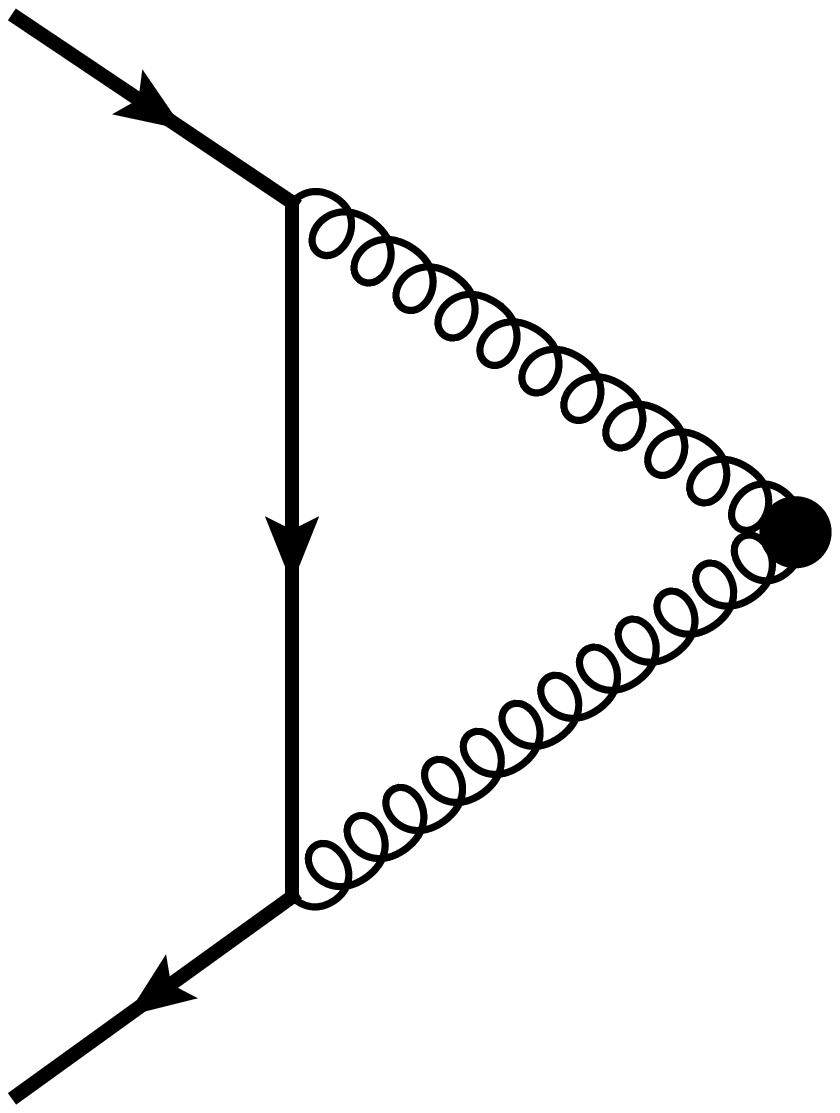}&
\hspace*{03mm}\includegraphics[width=1.5cm]{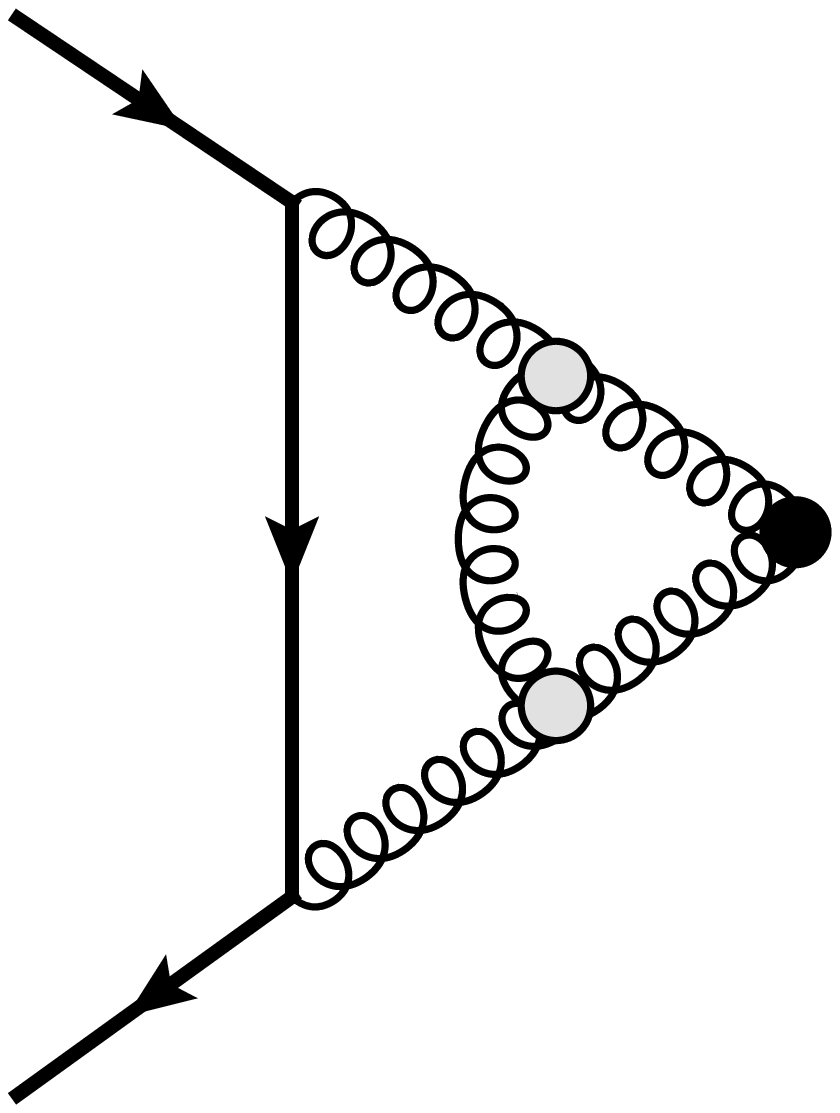}&
\hspace*{03mm}\includegraphics[width=1.8cm]{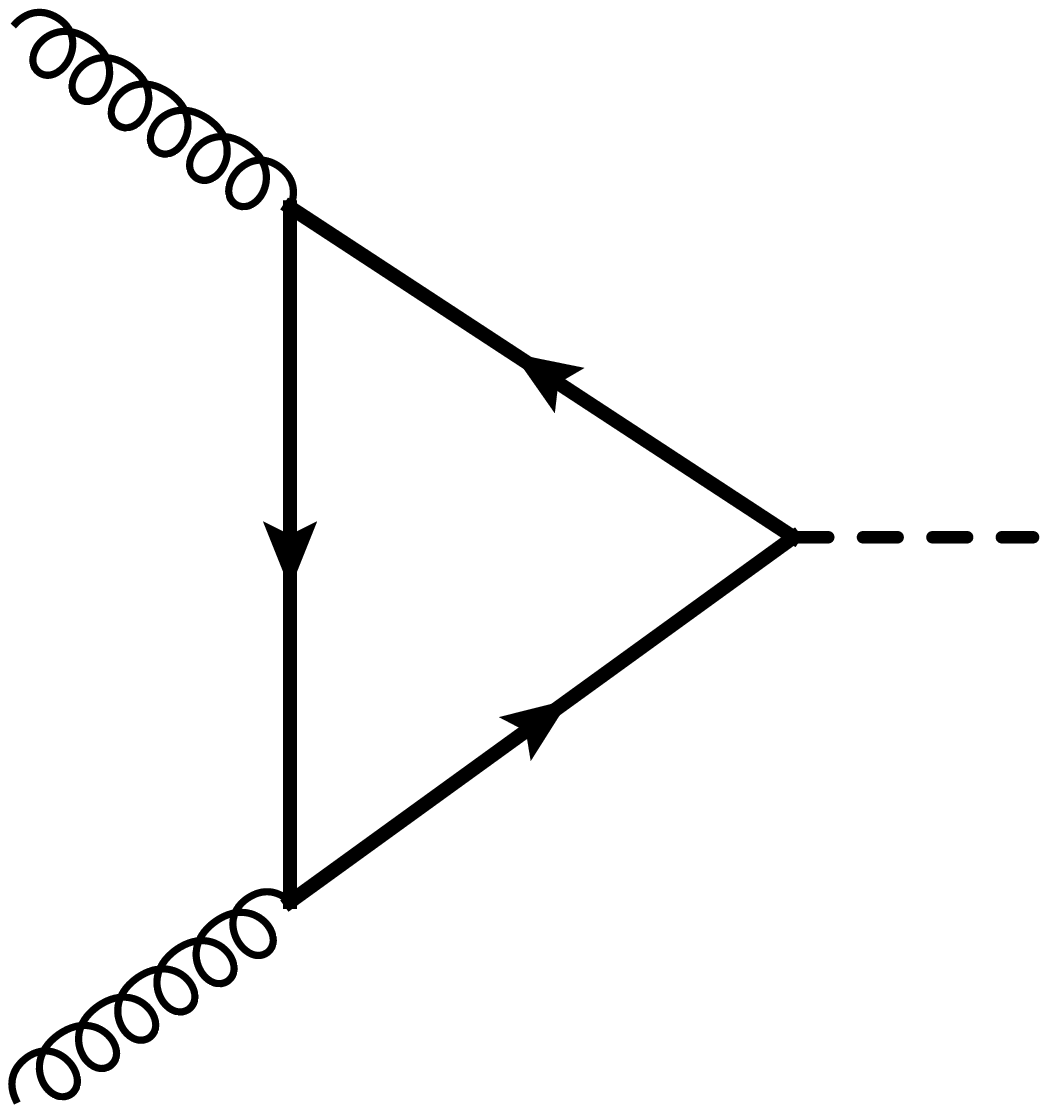}&
\hspace*{03mm}\includegraphics[width=1.8cm]{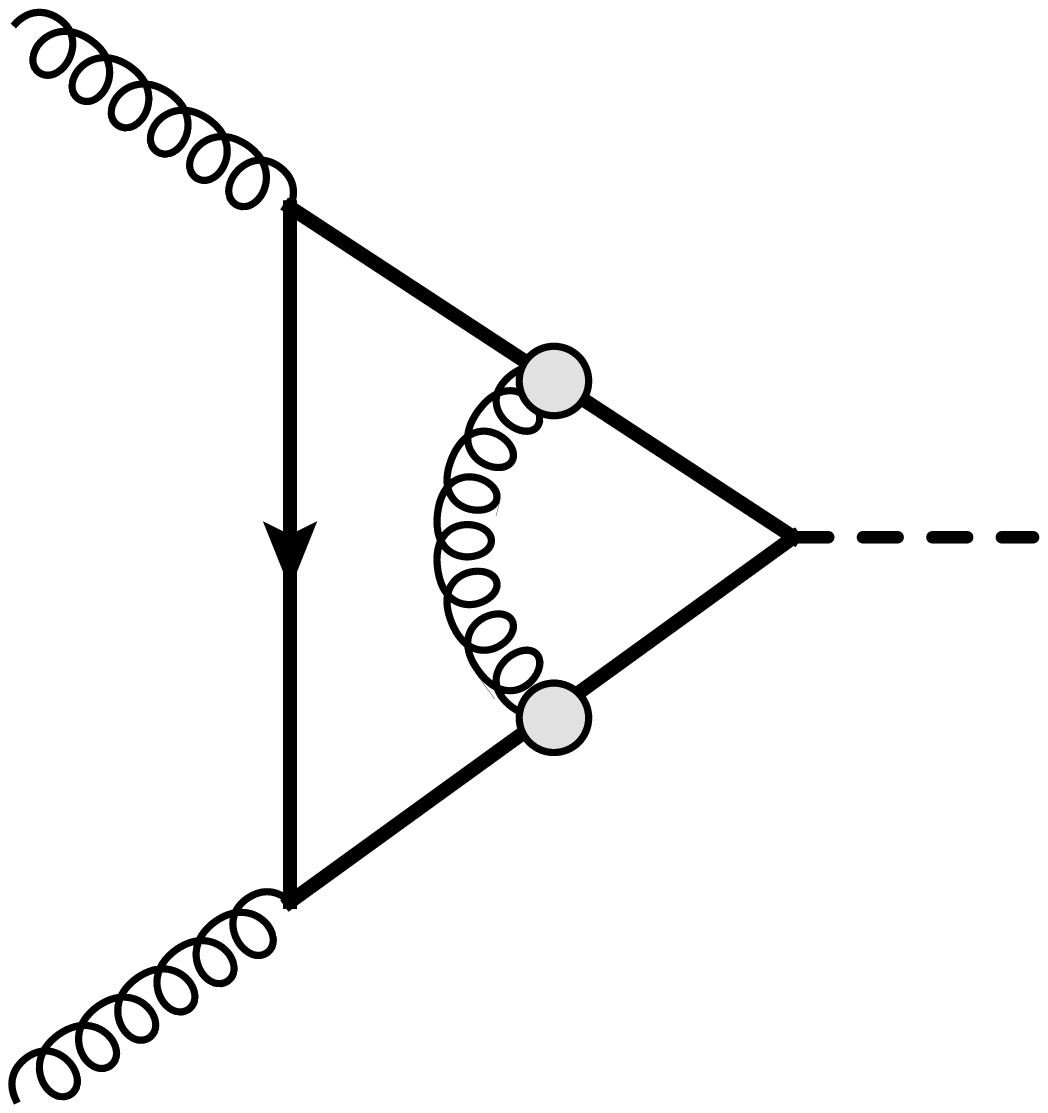}\\
(a)&\hspace*{03mm}(b)&\hspace*{03mm}(c)& \hspace*{03mm} (d)\\
\end{tabular}
\end{center}
\caption{\label{fig::1}  The leading order one-loop Feynman diagrams for  (a)
quark scattering by the $(G_{\mu\nu}^a)^2$ vertex (black circle) and (c) the
Higgs boson production in gluon fusion. The diagrams (b) and (d) with the
effective vertices (gray circles) defined in the text represent the
non-Sudakov double-logarithmic corrections to the process  (a) and (c),
respectively.}
\end{figure}

\begin{figure}[t]
\begin{center}
\begin{tabular}{cccc}
\includegraphics[width=1.5cm]{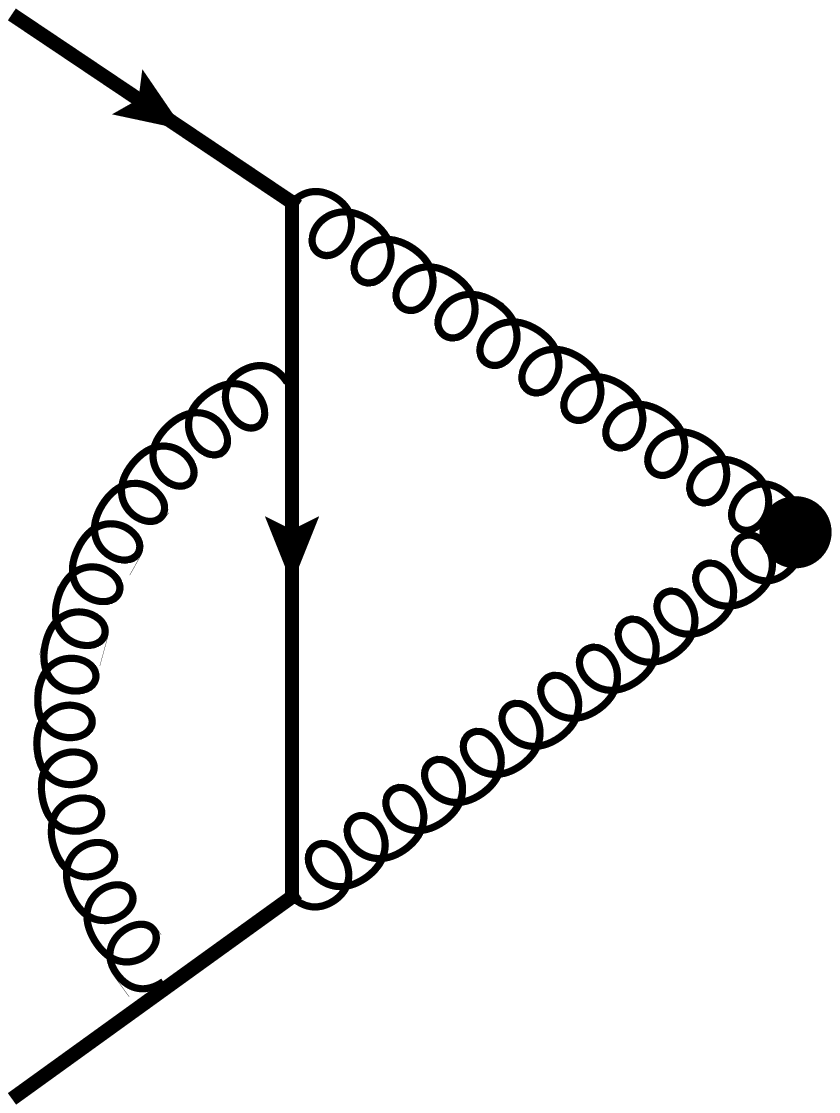}~\raisebox{9.5mm}{$\bfm{\to}$}&
\hspace*{00mm}\includegraphics[width=1.5cm]{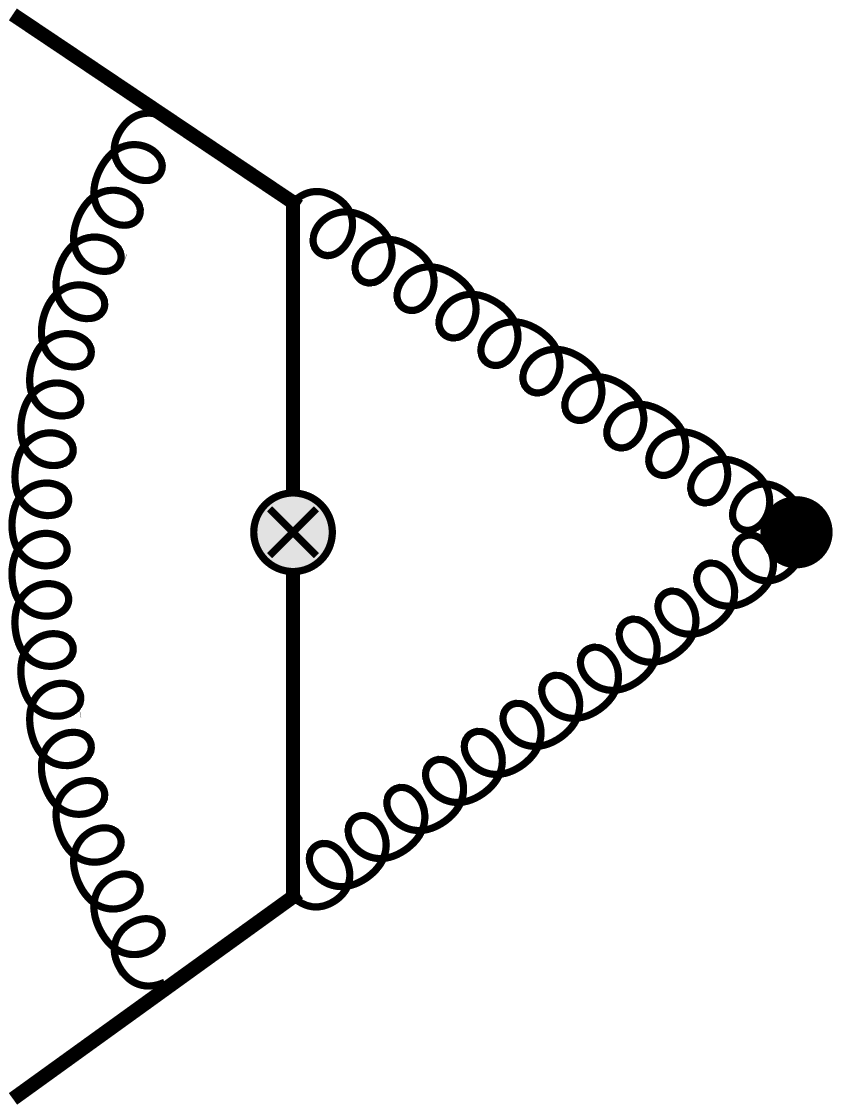}~\raisebox{9.5mm}{$\bfm{\to}$}&
\hspace*{00mm}\includegraphics[width=1.5cm]{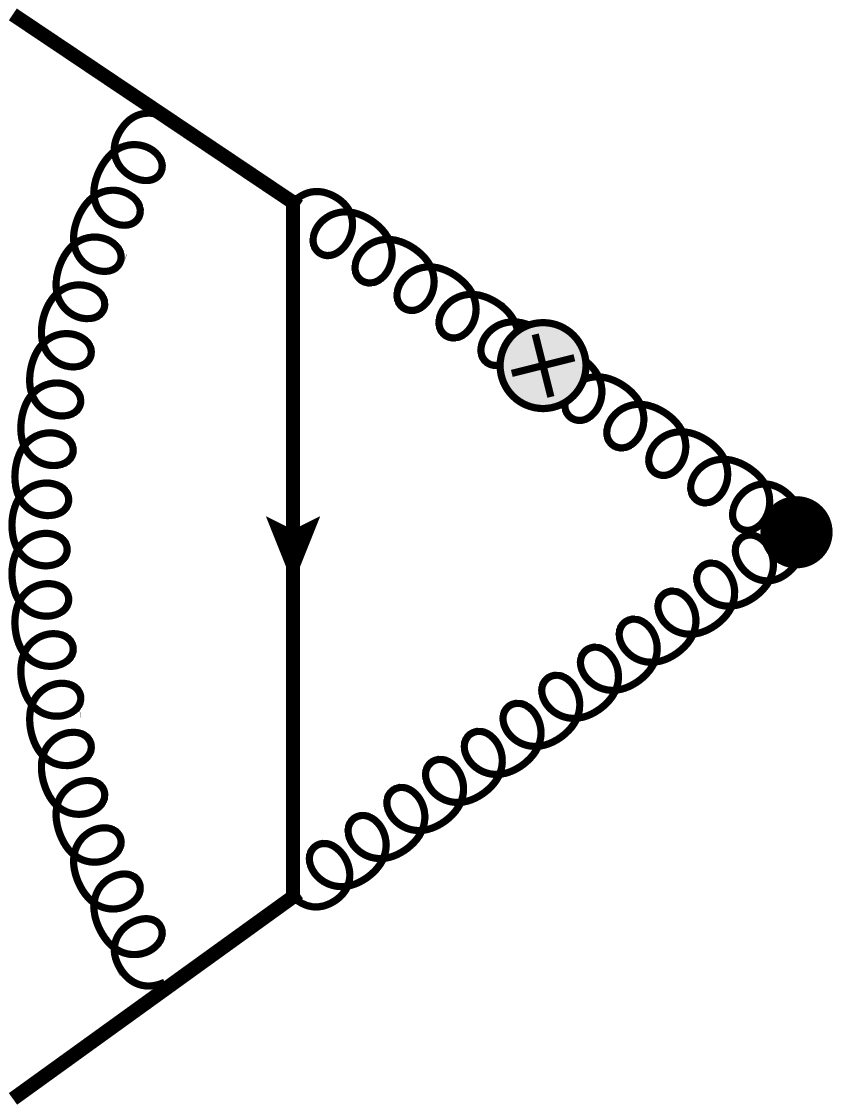}~\raisebox{9.5mm}{$\bfm{\to}$}&
\hspace*{00mm}\includegraphics[width=1.5cm]{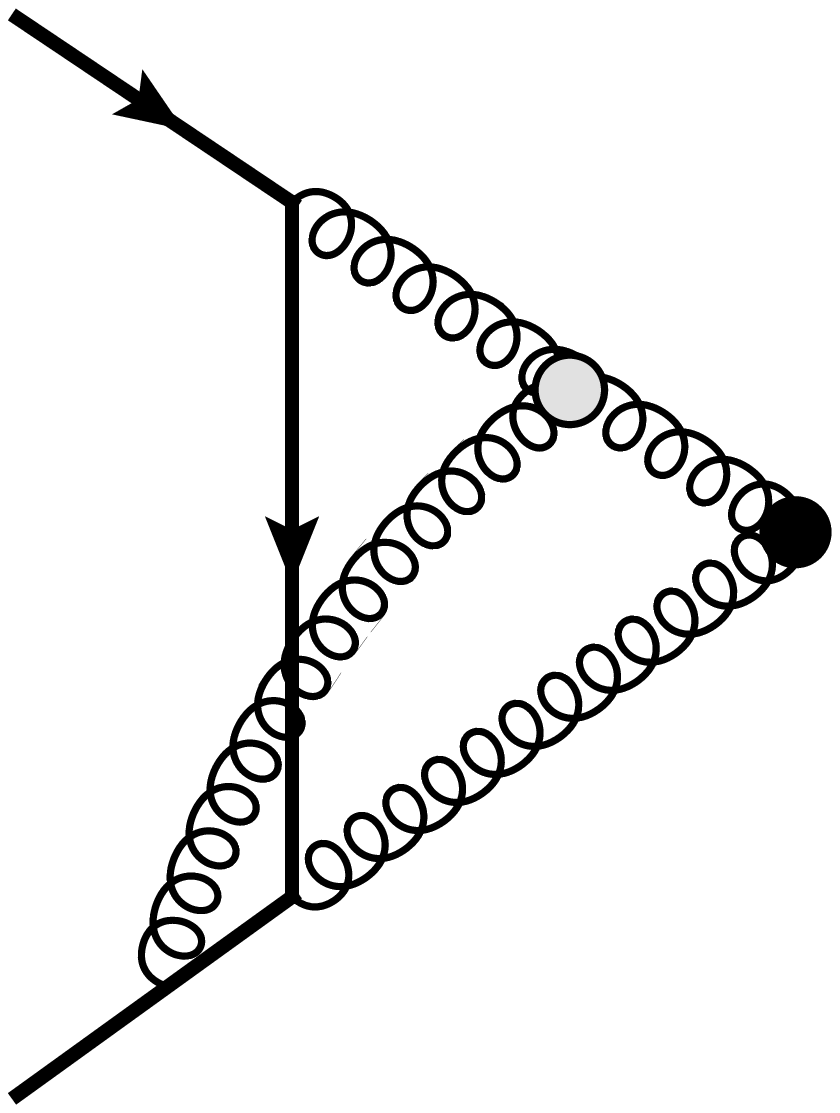}\\
(a)&\hspace*{00mm}(b)&\hspace*{00mm}(c)& \hspace*{00mm} (d)\\
\end{tabular}
\end{center}
\caption{\label{fig::2}  Diagramatic representation
of the sequence of identities which move the soft gauge boson vertex from the
soft quark to the eikonal gauge boson line, as explained in the text.}
\end{figure}

To introduce the main idea  of our approach we consider an amplitude ${\cal
G}$ for the  scattering of a quark of mass $m_q$, initial momentum $p_1$ and
final momentum $p_2$,  by a local operator $(G^a_{\mu\nu})^2$ of the gauge
field strength tensor. The origin of such a vertex is not relevant for our
discussion and one may suggest  that it describes the gluon field interaction
to the Higgs boson mediated by an infinitely heavy quark loop. We consider
the limit of the on-shell quark $p_1^2=p_2^2=m_q^2$ and the large Euclidean
momentum transfer $Q^2=-(p_2-p_1)^2$ when the ratio $\rho\equiv {m_q^2/Q^2}$
is positive and small.  In the light-cone coordinates $p_1\approx p_1^-$ and
$p_{2}\approx p_2^+$. The leading-order scattering is given by the one-loop
diagram in Fig.~\ref{fig::1}(a). Conservation of helicity  at high energy
requires a helicity flip on the virtual quark line. As a consequence at high
energy the amplitude  is suppressed  by the first power of $m_q$. The virtual
quark propagator then can be approximated  as follows $S(l)\approx{m_q\over
l^2-m_q^2}$. For $m_q\ll l\ll Q$ the  gauge boson propagators are eikonal
{\it i.e.} proportional to ${1\over 2p_il}$, and the diagram has  the
double-logarithmic scaling. Thus we have a typical situation when a soft
quark exchange generates the double-logarithmic contribution to the
mass-suppressed amplitude. As we see, the emission of the soft quark results
in the change of the  color group representation of a particle propagating
along the eikonal line, or  the {\it eikonal color charge nonconservation}.
This is a crucial feature of the process which plays an important role in
further analysis. Let us now consider the radiative corrections to the
amplitude and, for a moment, focus on an abelian case of the photon
interaction. Then in a covariant gauge the double-logarithmic corrections are
produced by a soft photon exchange between the external quark lines or an
external and the virtual quark lines. The key  idea of the approach is to
move the soft photon vertex from the virtual soft quark line to an eikonal
photon line through a sequence of identities graphically represented in
Fig.~\ref{fig::2}. Let us describe this sequence in more detail.
In a covariant gauge only $A^-$  light-cone component of the photon field can be
emitted by the eikonal quark line with the momentum $p_{2}$, while the emission
of the $A^+$ and transverse components is suppressed. Thus we can  use the
Ward identity to convert the diagram Fig.~\ref{fig::2}(a) into
Fig.~\ref{fig::2}(b) where the crossed circle on the quark propagator correspond
to the replacement $S(l)\to S(l)-S(l+l_g^+)$, with $l_g$ being the soft photon
momentum. By the momentum shift  $l\to l-l_g^+$ in the second term of the above
expression the crossed circle  can be moved to the eikonal line  which becomes
${1\over 2p_1l}-{1\over 2p_1(l+l_g^+)}$, Fig.~\ref{fig::2}(c). The opposite
eikonal line is not sensitive to this shift since $p_2^-\approx 0$. Finally by
using the inverted identity we  transform the diagram
Fig.~\ref{fig::2}(c) into Fig.~\ref{fig::2}(d) with an effective dipole coupling
$2e_q p_1^\mu$ to the {\it photon} line, where  $e_q$ is the {\it quark}
charge. Since  $p_1^+\approx 0$ we can replace $p_1l_g^+$  by $p_1 l_g$  in the
gauge boson propagator as long as $l_g\ll Q$ and after adding the symmetric
diagram we  get a structure characteristic  to the standard eikonal
factorization picture.  This factorization, however,  requires the summation
over all possible insertions of the soft photon vertex along each eikonal line
while in the case under consideration the diagram in Fig.~\ref{fig::1}(b) with
the soft exchange between the  photon  lines is missing. This diagram can be
added to complete the  factorization and then subtracted. Thus after factoring
out  the  soft photon exchange between the external quark lines the remaining
soft photon contribution is given by the diagram Fig.~\ref{fig::1}(b)  with the
coefficient  $-e_q^2$.  Note that the first Ward identity of the sequence  in
Fig.~\ref{fig::1}  is sufficient to prove the factorization of the soft photons
with the momentum $l_g\ll m_q$ as it has been done  in the original
paper~\cite{Yennie:1961ad}. This  algorithm however does not work for the
momentum interval $m_q\ll l_g\ll Q$ which does contribute to the
double-logarithmic  corrections. Our method extends the factorization to this
region at the expense of introducing the above subtraction term, which
compensates the charge variation of the eikonal line after the soft quark
emission.

The above result  can be generalized  to QCD in a straightforward way. The
difference with respect to the abelian case is that $e_q^2$  should be replaced
by the quadratic Casimir operator of the fundamental representation $C_F$  and
the contribution similar to Fig.~\ref{fig::1}(b) does exist in QCD  due to
gluon  self-coupling and is proportional to the quadratic Casimir operator the
adjoint  representation $C_A$.  Thus the part of the soft gluon exchange which
does not factorize into external lines is given by the diagram
Fig.~\ref{fig::1}(b) with the color weight $C_A-C_F$, which directly links it to
the variation of the color charge along the eikonal lines. We have verified the
above factorization by explicit evaluation of the two-loop  corrections in the
high-energy limit within the expansion by regions framework
\cite{Beneke:1997zp,Smirnov:1997gx,Smirnov:2002pj}. Since the emission of the
soft gluons from an eikonal line of a given color charge factorizes and
exponentiates \cite{Frenkel:1984pz} we can apply the above trick to an arbitrary
number of soft gluons. Hence the factorization formula for the double-logarithmic
corrections  to the amplitude becomes
\begin{equation}
{\cal G}= Z_{q}^2 g(-z){\cal G}^{(0)}\,,
\label{eq::Fq}
\end{equation}
where  ${\cal G}^{(0)}$ is the leading-order one-loop amplitude, $Z_{q}^2$ is
the standard Sudakov factor for a quark scattering, and the function $g(-z)$
incorporates the non-Sudakov contribution of Fig.~\ref{fig::1}(b) with an
arbitrary number of the effective  soft gluon exchanges.  The Sudakov factor
reads
\begin{equation}
 Z_{q}^2=\exp\left[-C_F\left({\alpha_s\over 2\pi}{\ln\rho\over \varepsilon}+x\right)\right],
\label{eq::Zq}
\end{equation}
where $x={\alpha_s\over 4\pi}\ln^2\!\!\rho$ is the double-logarithmic variable
and dimensional regularization with $d=4-2\varepsilon$  is used for  the
infrared divergences. The function $g(z)$ of the variable $z=(C_A-C_F)x$ is
normalized to $g(0)=1$ and can be obtained by the standard method
\cite{Penin:2014msa,Melnikov:2016emg,Liu:2017axv} in the form of the two-fold
integral
\begin{equation}
g(z)=2\int_0^1 {\rm d}\xi \int_{0}^{1-\xi}{\rm d}\eta e^{2z\eta\xi}
\label{eq::g}
\end{equation}
over the  normalized logarithmic variables $\eta=\ln v/\ln\rho $,
$\xi=\ln u/\ln\rho$  related to the Sudakov parametrization  of the  soft quark
momentum $l=up_1+vp_2+l_\perp$. The argument of the exponent in
Eq.~(\ref{eq::g}) corresponds to  the  single soft gluon contribution. The
integral Eq.~(\ref{eq::g}) can be solved in terms of  the  generalized
hypergeometric function
\begin{equation}
g(z)={}_2F_2\left(1,1;{3/2},2;{z/2}\right)=2\sum_0^\infty {n!\over (2n+2)!}(2z)^n
\label{eq::gseries}
\end{equation}
with the following asymptotic behavior at $z\to\infty$
\begin{equation}
g(-z)\sim {\ln(2z) +\gamma_E\over z}, \quad g(z)\sim \left({2\pi e^{z}\over z^{3}}\right)^{1/2}\!\!,
\label{eq::gasymp}
\end{equation}
where $\gamma_E=0.577215\ldots$ is the Euler constant. We have confirmed the
perturbative expansion of Eq.~(\ref{eq::Fq}) to ${\cal O}(\alpha_s^3)$ by
explicit evaluation of the three-loop double-logarithmic term adopting the
method of Ref.~\cite{Liu:2017axv}.

The above equations determine the amplitude  ${\cal G}$ in the  high-energy
limit in double-logarithmic approximation. Though this amplitude is of no
particular phenomenological interest, the result  can be used to find the
solution for the amplitude of Higgs boson production in gluon fusion mediated
by a bottom-quark loop mentioned in the introduction. Indeed, in the leading
order this  amplitude is given by the diagram in  Fig.~\ref{fig::1}(c). Since
the eikonal lines are characterized by the momentum and color charge but not
spin, the only difference  with respect to the previous case is the direction of
the color flow. Hence the  diagram in Fig.~\ref{fig::1}(d) which incorporates
the non-Sudakov part of the correction corresponds to the same function $g(z)$
with the opposite sign of the argument and the factorization formula takes the
form
\begin{equation}
{\cal M}^b_{gg\to H}= -Z^2_{g} g(z)
\left({3\over 2}\ln^2\!\!\rho\,\rho\right){\cal M}^{(0)}_{gg\to H}\,,
\label{eq::MH}
\end{equation}
where $\rho=m_b^2/m_h^2$ is now a Minkowskian parameter, the  heavy top-quark
loop mediated amplitude is used as the leading order approximation ${\cal
M}^{(0)}_{gg\to H}$ to have the mass suppression factor explicitly, and
\begin{equation}
 Z_{g}^2=\exp\left[{-{C_A\over\varepsilon^2}{\alpha_s\over 2\pi}}\right]
\label{eq::Zg}
 \end{equation}
is the Sudakov factor for  a gluon scattering. We have verified
Eq.~(\ref{eq::MH}) and, in particular, the  relation between the diagrams
imposed by the Ward identities by explicit two-loop calculation. The total
two-loop contribution also agrees with the analytical result  for the amplitude
with an arbitrary  value of the quark mass \cite{Anastasiou:2006hc} expanded in
the series in $\rho$.

\begin{figure}[t]
\begin{center}
\begin{tabular}{ccc}
\includegraphics[width=1.6cm]{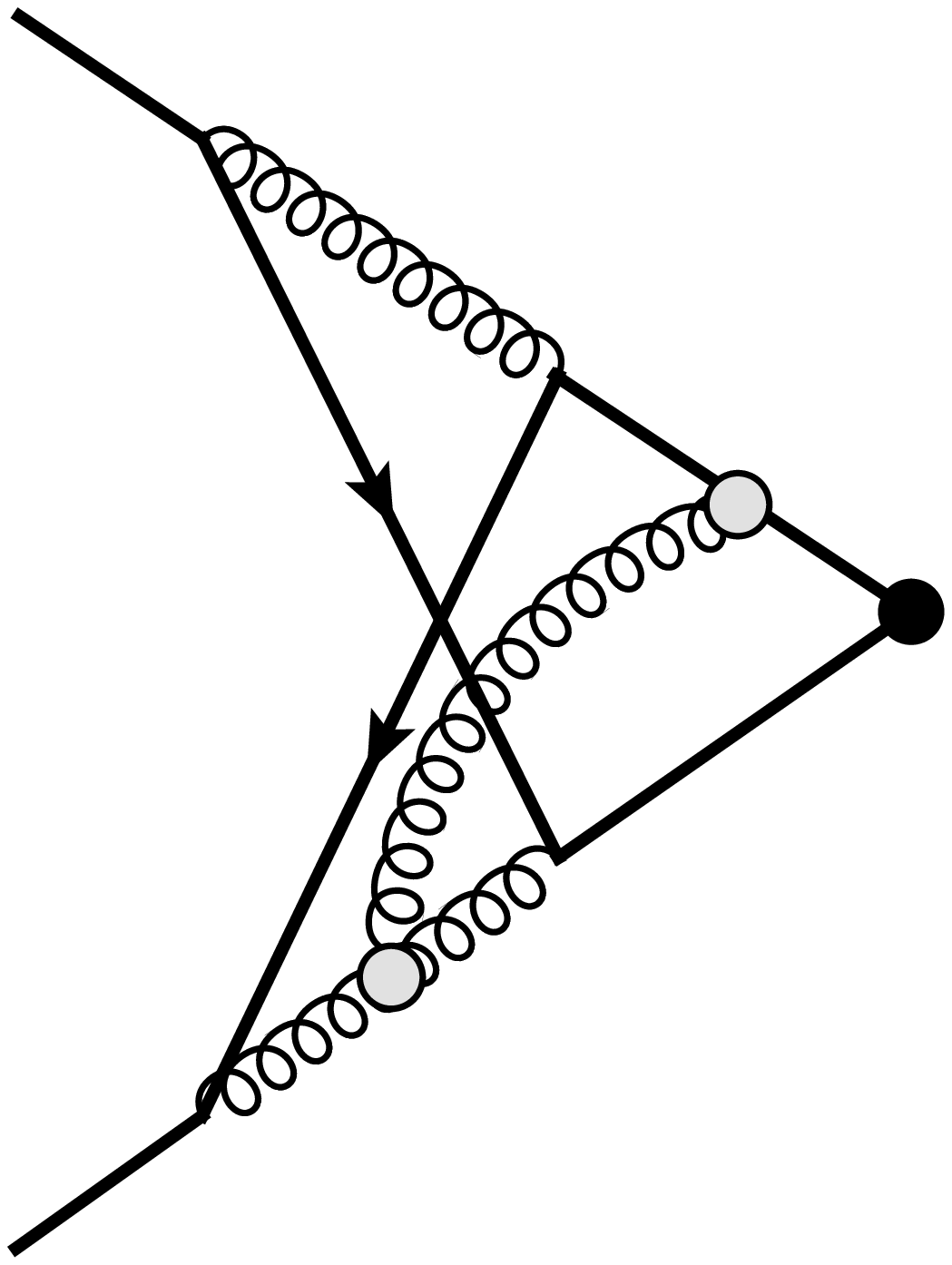}&
\hspace*{10mm}\includegraphics[width=1.5cm]{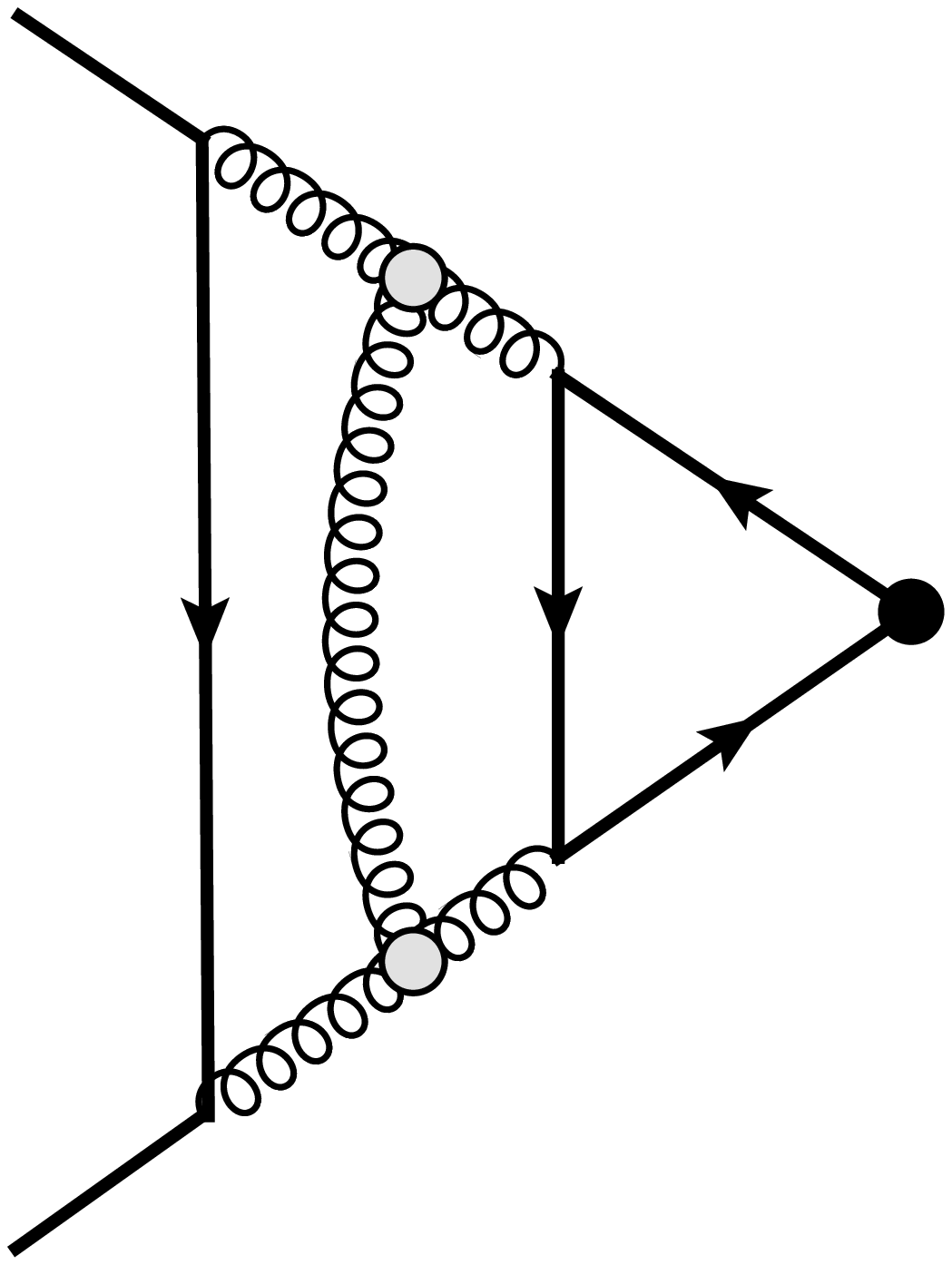}&
\hspace*{10mm}\includegraphics[width=1.5cm]{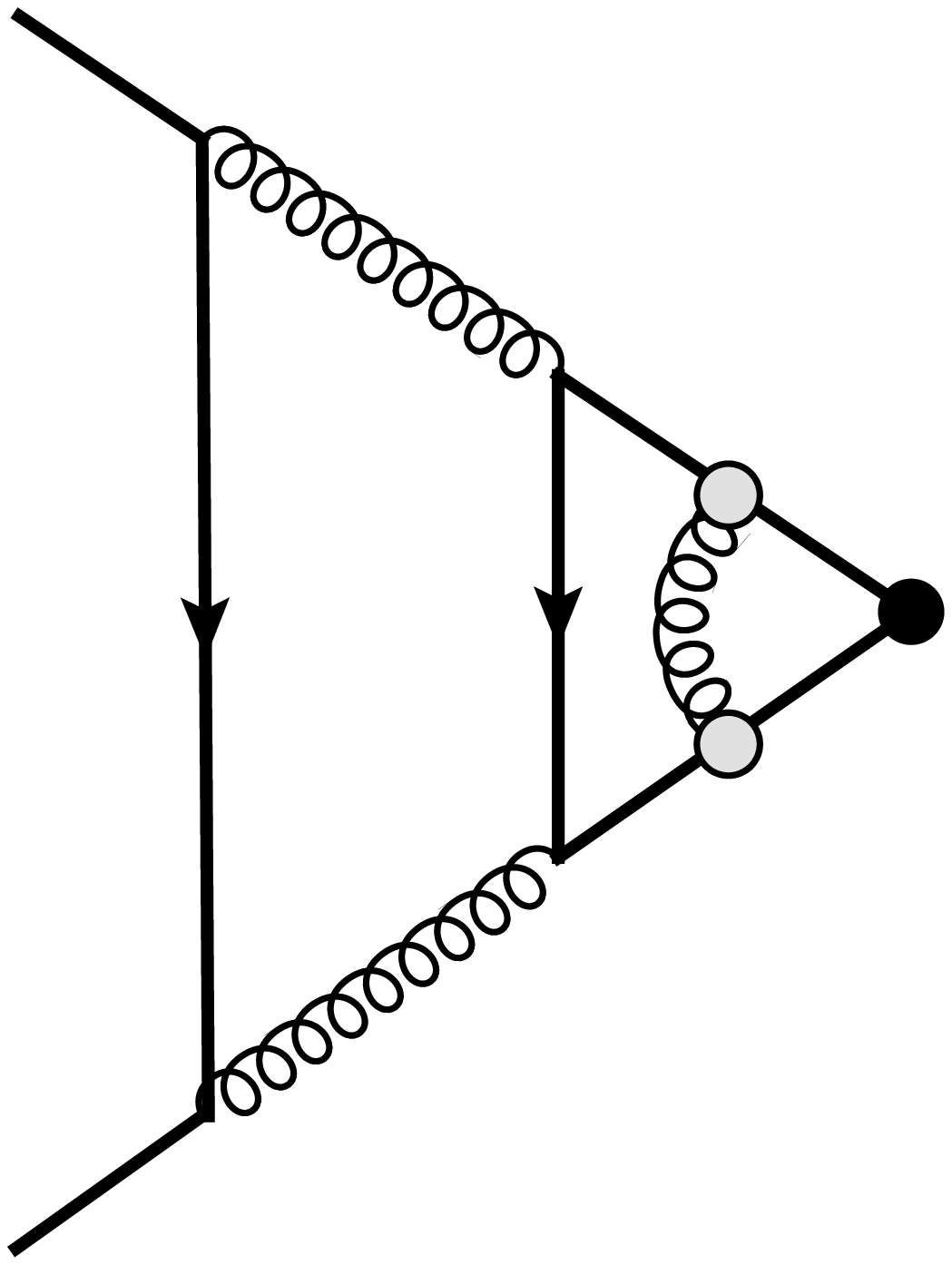}\\
(a)&\hspace*{10mm}(b)&\hspace*{10mm}(c)\\
\end{tabular}
\end{center}
\caption{\label{fig::3} The diagrams with an effective soft gluon exchange
which incorporate the non-Sudakov double-logarithmic corrections to
(a) vector and (b,c) scalar form factor  of a quark. The symmetric diagrams
are not shown.}
\end{figure}

Let us now consider a more complex problem of finding the asymptotic behavior of
the leading mass-suppressed  contribution to the amplitude of quark scattering in
an external field. We start with the vector field case.  At high energy
and in the double-logarithmic approximation the  deviation of the corresponding
amplitude from the  Born approximation is described by the Dirac form-factor
$F_1$. Its asymptotic expansion can be written as follows
\begin{equation}
F_1=Z_{q}^2\sum_n \rho^n F^{(n)}_1,
\label{eq::F1series}
\end{equation}
where $F^{(n)}_1$ are given by the power series in $\alpha_s$ with the
coefficients depending on $\rho$ only logarithmically.  Since the Sudakov
corrections in Eq.~(\ref{eq::F1series}) are factored out, in the
double-logarithmic approximation the leading term of  the expansion   is just
$F_1^{(0)}=1$. The double-logarithmic corrections  to the leading
power-suppressed term $F_1^{(1)}$ are induced by the nonplanar soft quark pair
exchange and start with the two-loop contribution
\cite{Penin:2014msa,Penin:2016wiw,Liu:2017axv}. Following the algorithm
described above we reduce the non-Sudakov part of the corrections to the diagram
in Fig.~\ref{fig::3}(a), and the symmetric one. In these diagrams, as in
Fig.~\ref{fig::1}(b), the effective soft gluon exchange with the color weight
$C_A-C_F$ connects the eikonal lines with a given soft quark momentum.  Note
that in the double-logarithmic region the light-cone components of the soft
quark momenta are ordered along the eikonal lines so that in each eikonal
propagator only one  momentum should  be retained, which  determines the position
of the effective vertices in Fig.~\ref{fig::3}(a). The corresponding
factorization formula  for the leading power-suppressed term reads
\begin{equation}
F_1^{(1)}={C_F(C_A-2C_F)\over 6}x^2f(-z)\,,
\label{eq::F1result}
\end{equation}
where the function $f(-z)$ incorporates the non-Sudakov contribution of
Fig.~\ref{fig::3}(a) with an arbitrary number of the effective soft gluon
exchanges and is normalized to the two-loop result $f(0)=1$.  This function has
an integral  representation similar to Eq.~(\ref{eq::g})
\begin{eqnarray}
f(z)&=&12\int_0^1{\rm d}\eta_1\!\!\int_{\eta_1}^{1}{\rm d}\eta_2
\!\!\int_0^{1-\eta_2}\!\!\!\!\!\!\!\!{\rm d}\xi_2\!\!\int_{\xi_2}^{1-\eta_1}
\!\!\!\!\!\!\!\! {\rm d}\xi_1
\,e^{2z\eta_1(\xi_1-\xi_2)}
\nonumber\\
&&\times e^{2z\xi_2(\eta_2-\eta_1)}
\,,
\label{eq::f}
\end{eqnarray}
where the integration is performed over the logarithmic Sudakov variables
for each soft quark momenta and the exponential factors correspond to
Fig.~\ref{fig::3}(a) and  the symmetric diagram. We are not
able to solve the  four-fold integral Eq.~(\ref{eq::f}) in a closed
analytic form. However,  the coefficients of the series
$f(z)=1+\sum_{n=1}^\infty c_n z^n $ can be computed for any given $n$
corresponding to the $(n+2)$-loop double-logarithmic contribution and have the
following large-$n$ behavior  $c_n\sim {\ln n\over n! 2^{n} n^{5/2}}$. The first
seven coefficients of the series are listed in Table~\ref{tab::cn}.  The asymptotic
behavior of the function at $z\to \infty$ reads
\begin{equation}
f(-z)\sim   C_-\left({\ln{z} \over z}\right)^2\!,
\quad f(z)\sim C_+\ln z\left({e^{z}\over z^5}\right)^{1/2}\!\!,
\label{eq::fasym}
\end{equation}
where the constant $C_-=3.6\ldots$, $C_+=14.8\ldots$ are found numerically.

\begin{table}[t]
  \begin{ruledtabular}
    \begin{tabular}{c|c|c|c|c|c|c|c}
      $n$ & $1$ & $2$ & $3$ & $4$ & $5$ & $6$ & $7$
      \\
      \hline
      \hline
      $2^nn^2n!c_n$ &  ${2\over 5}$ & ${88\over  105}$ & ${8\over 7}$
      & ${70144\over 51975}$ & ${640\over 429}$ &
      ${25344\over 15925}$ & ${2727424\over 1640925}$
      \\
    \end{tabular}
    \end{ruledtabular}
    \caption{\label{tab::cn}
      The   normalized coefficients of the Taylor series for the function
      $f(z)$, Eq.~(\ref{eq::f}), up to  $n=7$.}
\end{table}

Let us now consider quark scattering  in the external scalar field
parametrized by the scalar form factor $F_S$. In the  equivalent notations
for the leading power-suppressed term $F_S^{(1)}$  we obtain
\begin{equation}
F_S^{(1)}=-{C_FT_F\over 3}x^2f_S(-z)\,,
\label{eq::FSresult}
\end{equation}
where $T_F=1/2$ and the function
\begin{equation}
f_S(z)=24\int_0^1{\rm d}\eta_1\!\!\int_{0}^{1-\eta_1}\!\!\!\!\!\!\!\!{\rm d}\xi_1
\!\!\int_{\eta_1}^{1-\xi_1}\!\!\!\!\!\!\!\!{\rm d}\eta_2\!\!\int_{\xi_1}^{1-\eta_2}
\!\!\!\!\!\!\!\! {\rm d}\xi_2
\,e^{2z\eta_2\xi_2}e^{-2z\eta_1\xi_1}
\label{eq::fS}
\end{equation}
is determined by the planar diagrams in Fig.~\ref{fig::3}(b,c) with the
corresponding exponential factors given separately. Amazingly, though the
topology of the diagrams in Fig.~\ref{fig::3}(a) and  Fig.~\ref{fig::3}(b,c)
is completely different, Eqs.~(\ref{eq::f}) and (\ref{eq::fS}) describe  the
{\it same function}  $f_S(z)\equiv f(z)$ as it can be easily verified. For
the moment we do not have a plausible explanation of this universality.  At
the same time it is straightforward to extend the analysis to the axial $F_A$
and the pseudoscalar $F_P$ form factors, for which we obtain the result in
the form of Eq.~(\ref{eq::F1result}) and Eq.~(\ref{eq::FSresult}) with
$f_A(z)=-f(z)$ and $f_P(z)=f(z)$, respectively.

Our results   agree with the asymptotic expansion of the exact two-loop
expressions for the form factors
\cite{Bernreuther:2004ih,Bernreuther:2004th,Bernreuther:2005gw}. The result
for the vector form factor agrees with the explicit calculation of the
three-loop  double-logarithmic term \cite{Liu:2017axv} including the
relations between the diagrams imposed by the Ward identities.  For $C_A=0$
and $C_F=1$  Eq.~(\ref{eq::F1result}) agrees with the all-order QED result
\cite{Penin:2014msa}.\footnote{In Ref.~\cite{Penin:2014msa} only the singular
part of the Sudakov corrections Eq.~(\ref{eq::Zq}) has been factored out
while $e^{-x}$ part has been absorbed into the definition of $F_1^{(1)}$.
Such a decomposition however is not physical since the soft real emission
cancels all the double logarithmic terms in Eq.~(\ref{eq::Zq}) rather than
its singular part.}

Thus we have performed the first systematic analysis of the high-energy
asymptotic behaviour of the  QCD amplitudes beyond the leading-power
approximation and  derived  all-order double-logarithmic result  for the
leading mass-suppressed terms in typical two-scale problems. After separating
the standard Sudakov factors the remaining non-Sudakov double-logarithmic
corrections are described by two universal functions $g(\pm z)$ and $f(\pm z)$,
Eqs.~(\ref{eq::gseries}) and (\ref{eq::f}), of the variable $z={\alpha_s\over
4\pi}(C_A-C_F)\ln^2(m_q^2/Q^2)$ for the processes with single and double soft
quark exchange, respectively.  Note that in general the amplitudes with
larger number of scattering  particles, such as Bhabha scattering in QED
\cite{Penin:2016wiw}, get contributions from both single and double soft
fermion exchange and the corresponding asymptotic expressions involve both
functions $g(z)$ and $f(z)$. These functions play the role of ``Sudakov
exponent'' for the non-Sudakov double-logarithmic corrections. They are
exponentially {\it  enhanced} for large positive values of the argument and
power suppressed for the large negative values. Our analysis reveals a
nontrivial relation  between the asymptotic behavior of different amplitudes
and the amplitudes in different gauge theories.  In particular, it
demonstrates  that if a QCD amplitude gets the exponential enhancement at
high energy, the same amplitude in QED is  suppressed by a power of the large
logarithm,  and {\it vice versa}. Note that the enhancement of the
mass-suppressed amplitudes by the double-logarithmic corrections has been
already observed in QED \cite{Penin:2016wiw,Gorshkov:1966ht}. The method
introduced above is quite general and can be used for the analysis of the
more  complex processes and observables such as Higgs boson production in
association with a jet  and the  Higgs boson  transverse momentum
distribution, which however is beyond the scope of this Letter.

\begin{acknowledgements}
A.P.  would like to thank  Kirill Melnikov for numerous  discussions and
Babis Anastasiou for useful communication.
The work of A.P. is supported in part by NSERC and  Perimeter Institute for
Theoretical Physics. The work of T.L. is supported by NSERC.
\end{acknowledgements}


\end{document}